\documentclass[aps,superscriptaddress,preprintnumbers,tightenlines,nofootinbib,11pt]{revtex4-1}

\usepackage[english]{babel}
\usepackage{amsmath,amssymb,amsbsy,amstext,amsthm,booktabs,simplewick,exscale,relsize,slashed,graphicx,amsfonts,upgreek}
\usepackage{tabu,multirow,color,dcolumn,bm,enumerate}
\usepackage{latexsym}
\usepackage{mathrsfs}
\usepackage{amsthm}
\usepackage{outlines}

\def\beq{\begin{eqnarray}}
\def\eeq{\end{eqnarray}}
\def\bea{\begin{eqnarray}}
\def\eea{\end{eqnarray}}

\def\tev{\, {\rm TeV}}
\def\gev{\, {\rm GeV}}

\def\fm{\, {\rm fm}}

\def\fb{\, {\rm fb}}
\def\ab{\, {\rm ab}}
\newcommand{\gsim}{\lower.7ex\hbox{$\;\stackrel{\textstyle>}{\sim}\;$}}
\newcommand{\lsim}{\lower.7ex\hbox{$\;\stackrel{\textstyle<}{\sim}\;$}}

\def\pb{\, {\rm pb}}
\def\s{\, {\rm s}}

\def\yr{\, {\rm yrs}}

\usepackage{setspace}

\begin{document}

\preprint{UH-511-1240-2014}

\title{Proton annihilation at hadron colliders and Kamioka: \\ high-energy versus high-luminosity}

\author{Joseph Bramante\footnote{\tt jbraman2@nd.edu}}
\affiliation{Department of Physics, University of Notre Dame, 225 Nieuwland Hall, Notre Dame IN, 46556, USA}
\author{Jason Kumar\footnote{\tt jkumar@hawaii.edu}}
\affiliation{Department of Physics and Astronomy, University of Hawaii at Manoa, 2505 Correa Road, Honolulu HI, 96822, USA}
\author{John Learned\footnote{\tt jgl@phys.hawaii.edu}}
\affiliation{Department of Physics and Astronomy, University of Hawaii at Manoa, 2505 Correa Road, Honolulu HI, 96822, USA}

\begin{abstract}
We examine models and prospects for proton annihilation to dileptons, a process which violates baryon and lepton number each by two. We determine that currently Super-Kamiokande would place the most draconian bound on $pp \rightarrow \ell^+ \ell^+$, ruling out new physics below a scale of $\sim 1.6~\tev$. We also find present and future hadron collider sensitivity to these processes. While 8 TeV LHC data excludes new physics at a scale below $\sim 800~\gev$, the reach of a 14 TeV LHC run is $\sim 1.8~\tev$, putting it on par with the sensitivity of Super-Kamiokande. On the other hand, a 100 TeV proton-proton collider would be sensitive to proton annihilation at a scale up to 10 TeV, allowing it to far exceed the reach of both Super-Kamiokande and the projected 2 TeV reach of Hyper-Kamiokande. Constraints from neutron star observation and cosmological evolution are not competitive.
Therefore, although high-luminosity water Cherenkov experiments currently place the leading bounds on baryon and lepton number violation, next generation high-energy hadron colliders will begin surpassing them in sensitivity to some $B/L$-violating processes.
\end{abstract}

\maketitle

\tableofcontents
\section{Introduction}
While commendable progress has been made in understanding the interactions, masses, and charges of Standard Model particles, the source of the universe's baryon asymmetry remains at large. Planck and WMAP satellite measurements of the cosmic microwave background set the baryon-to-photon number density ratio~\cite{Ade:2013zuv,Hinshaw:2012aka}, at
\begin{align}
n_B / n_\gamma  = \left( 6.1 \pm 0.2 \right) \times 10^{-10},
\end{align}
whereas the expected ratio in a universe without a baryon asymmetry is eight orders of magnitude smaller. Although sources of baryon number, $C$ and $CP$ violation are required to explain this discrepancy~\cite{Sakharov:1967dj}, there is no detailed understanding of how this is achieved in nature.  $B+L$ charge is violated by quantum anomalies in the Standard Model (see e.g. Refs.~\cite{Kuzmin:1985mm,Dine:2003ax,Buchmuller:2005eh}), but this symmetry may be restored by the effects of new physics beyond the Standard Model (BSM), leaving all possibilities open.

A related question is whether or not the proton is exactly stable.
The observed hyperstability of the proton
has prompted models with baryon-stabilizing symmetries which extend deep into the
UV~\cite{Arnold:2009ay,Dulaney:2010dj,FileviezPerez:2011dg,FileviezPerez:2011pt,Baldes:2011mh,Csaki:2011ge,Graham:2012th,Arnold:2012sd,Patel:2013zla,Arnold:2013cva,Barry:2013nva,Perez:2013tea,Graham:2014vya,Perez:2014qfa,Faroughy:2014tfa}. $B$-violating processes which are introduced
to explain the baryon asymmetry can also destabilize the proton, contradicting the extremely tight bounds on the proton lifetime
set by Super-Kamiokande~\cite{Nishino:2009aa},
\bea
T_{p} > 8.2 \times 10^{33} ~{\rm yrs}.
\eea
A process which permits $\Delta (B-L)=0$, $\Delta (B+L)=\pm 2$
could allow the generation of a baryon asymmetry, but might also allow the process $p \rightarrow e^+ X$, and the rate of this process is constrained at late cosmological times by the observed relic abundance of protons. In this regard, it is worth noting that the electroweak sphaleron processes,
which contribute to the baryon asymmetry within the electroweak baryogenesis or leptogenesis frameworks,
are consistent with the current abundance of protons because they are exponentially suppressed at temperatures below the weak scale.

The abundance of protons and the existence of a baryon asymmetry in the current epoch are
often related to the assumption that protons cannot self-annihilate.  The essential logic is
that the lightest particle
charged under an exact continuous symmetry cannot self-annihilate, because there exists no
other possible final state with the same charge which is kinematically accessible.  But although
the proton is charged under the exact symmetry $U(1)_{EM}$, there are several lighter states with
the same charge ($e^+$, $\pi^+$, $K^+$).  The proton is also the lightest particle with non-zero
baryon number, but the associated symmetry would only forbid self-annihilation if it were an exact continuous
symmetry; if baryon number were instead the charge of an exact discrete $Z_2$ symmetry, then proton
annihilation would be perfectly consistent, although the proton would be exactly stable.
The continued existence of a baryon asymmetry requires only that the $pp$ annihilation cross section be small
enough that the annihilation process froze out in the early universe before depleting the baryon asymmetry; this
constraint would be easily satisfied provided $\langle \sigma_A^{pp} v \rangle \ll 1~\pb$.  If this condition is
satisfied, then proton annihilation would be frozen out well before Big Bang Nucleosynthesis (BBN), implying that it would have a
negligible effect on BBN and all subsequent cosmological evolution.

There have been a variety of studies of scenarios in which baryon number can be violated even though the proton is exactly stable~\cite{FileviezPerez:2011dg}.
But there are some interesting distinctions between proton-proton annihilation and the well-studied example of $n-\bar n$ oscillation, wherein
a neutron oscillates into its anti-particle, which then annihilates against another neutron.  One can
describe this scenario equivalently with an $n-\bar n$ mixing term in the mass matrix.  From this perspective, the true mass eigenstates are real fermions which are linear combinations of $n$ and $\bar n$ states and have a very small mass splitting, and the particles which annihilate are indistinguishable from their anti-particles.
We contrast this with proton-proton annihilation, a process which can only be interpreted as particle self-annihilation
because the anti-particle is distinguishable from the particle.

Indeed, the proton annihilation process $pp \rightarrow K^+ K^+$ ($\Delta B=\pm 2$, $\Delta L=0$) has been previously considered theoretically and
experimentally, motivated by studies of R-parity violating MSSM models
(the process $pp \rightarrow \pi^+ \pi^+$ is similar, but is not motivated by SUSY constructions).
In this study we focus on $pp \to \ell^+ \ell^+$ ($\Delta (B+L) = \pm 4$) and do not assume a particular fundamental theory, but instead present simplified models and $B+L$ violating effective operators which are allowed by the residual symmetries, and  which are obtained after integrating
out whatever heavy particles arise from the underlying UV completion.
Note that $pp \to \ell^+ \ell^+$ is the simplest proton annihilation process in a theory where the proton is stable and $B-L$ is conserved.
As colliders and experiments with large fiducial volumes gain sensitivity to higher energies and luminosities, $B+L$ violating operators merit attention as possible portals to primordial particle asymmetries.\footnote{A timely result~\cite{Perez:2014fja} of particular import to this study has shown that $B+L$ violating operators can induce CP violating interactions through the electroweak vacuum angle.}
We will find that there are two relevant effective field theories, a quark-level
effective field theory relevant for high-energy collider processes and a hadron-level effective field theory relevant for the low-energy
processes which occur at high-luminosity rare event experiments.

\begin{figure}[h!]
\begin{center}
\includegraphics[scale=1]{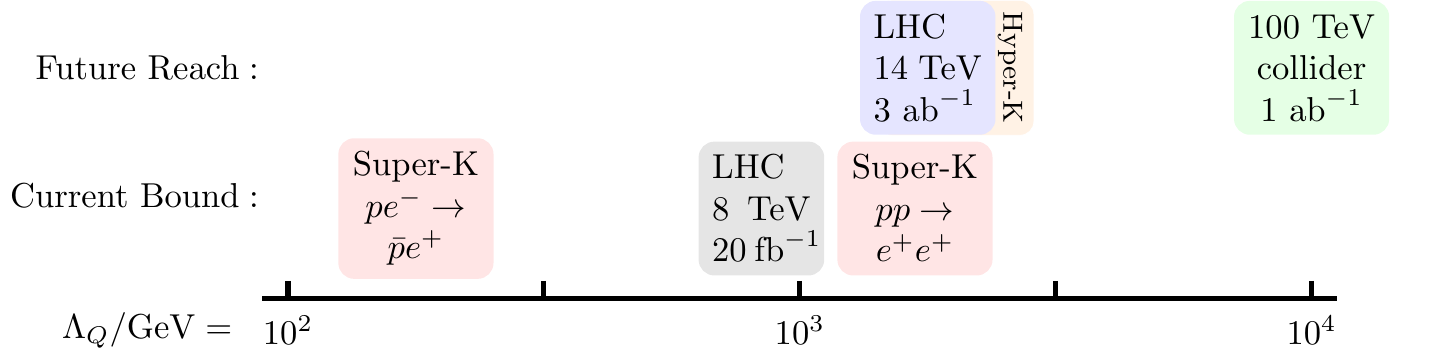}
\end{center}
\caption{This schematic shows the present and future sensitivity to diproton-dilepton coupling. While Super-K has set the best present limits, a future 100 TeV hadron collider could be the first to uncover diproton-dilepton couplings. $\Lambda_Q$ roughly sets the scale of new physics. More precisely, the cutoff shown ($\Lambda_Q$) is the cutoff of a dimension-12 operator coupling dileptons to diprotons through quarks, although it can be related to the dimension-6 operator (no quarks) cutoff via $\Lambda_{Q}^4 \simeq \Lambda ~\rm{(0.22 ~GeV)^3} $ (see Section \ref{sec:loweproc} and Eq. \eqref{eq:matching}). }
\label{fig:boundline}
\end{figure}

This article explores bounds on and prospects for low energy process with $\Delta (B+L) = \pm 4$, specifically protons annihilating to and with leptons, $p^+ p^+ \to \ell^+ \ell^+$, $p^+ e^- \to p^- e^+$, and associated processes related by a weak isospin transformation, $n n \to \bar \nu \bar \nu$.
We relate these low energy constraints to bounds on dimension twelve operators which foment processes like $\bar{u} \bar{u} \to u  u  d  d  e^- e^-$ at high-energy colliders, and determine what prospects lie ahead for $\Delta (B+L) = \pm 4$ processes at the large hadron colliders. Although Super and Hyper-Kamiokande will set the most stringent bounds on $\Delta (B+L) = \pm 4$ operators, a future 100 TeV hadron collider would offer unparalleled access to these processes, as we show in Figure \ref{fig:boundline}.
Indeed, we will show that this process is more amenable to collider searches than processes from lower dimensional operators, such as $pp \rightarrow \pi^+ \pi^+$.

In Section \ref{sec:loweproc} we present models for proton annihilation to same sign dileptons, and define effective operators at both quark level (dimension 12) and hadron level
(dimension 6) that violate both baryon and lepton number by two. In Section \ref{sec:superk} we calculate bounds for these operators by reinterpreting
Super-Kamiokande limits on $n-\bar n$ oscillation and proton decay. In Section \ref{sec:nstars} we demonstrate that neutron star formation enforces a
mild bound on these baryon and lepton number violating processes. The ability of present and future hadron colliders to uncover
dimension-12 dinucleon-dilepton couplings is examined in Section \ref{sec:hadroncolliders}, which recasts an LHC search
for $\ell^\pm \ell^\pm + 4 ~\rm{jets}$ events. We conclude in Section \ref{sec:conclusion} by emphasizing the importance of a 100 TeV
collider, which would be the first collider to probe dinucleon annihilation to dileptons at mediator energies surpassing those of
existing and planned water Cherenkov detectors.

\section{Low energy proton annihilation to positrons}
\label{sec:loweproc}

It is easiest to consider proton annihilation processes from the standpoint of the global symmetry group $U(1)_B \times U(1)_L$. This symmetry group
can be broken entirely, or to a subgroup which may contain discrete factors. Any process involving color singlets which conserves electromagnetic charge
and respects the remaining unbroken global symmetry group can be expected to have a non-vanishing amplitude.
A table of unbroken symmetry groups, along with possible proton decay processes, and alternative signatures if
proton decay is forbidden, are listed in Table~\ref{Table:Symmetries}.
We will focus on an unbroken a $Z_4^{B+L} \times U(1)_{B-L}$ global symmetry group which stabilizes the
proton, but permits proton annihilation.

\renewcommand\arraystretch{1.2}
\begin{table}
\begin{tabular}{|l|l|l|}
  \hline
  symmetries & proton decay & alternative signature \\
  \hline
  $U(1)_{B-L} \times Z_2^{B+L}$ & $p \rightarrow e^+ X$ & --- \\
  $U(1)_L$ & $p \rightarrow \pi^+ X$ & --- \\
  $Z_2^B \times U(1)_L$ & --- & $n -\bar n$ oscillation; $pp \rightarrow \pi^+ \pi^+, K^+ K^+$ \\
  $Z_4^{B+L} \times U(1)_{B-L}$ & --- & $ pp \rightarrow \ell^+ \ell^+ $   \\
  \hline
\end{tabular}
\caption{Some possible unbroken subgroups of $U(1)_B \times U(1)_L$, along with allowed proton decay
channels and alternative signatures (if proton decay is forbidden). The state $X$ can consist of Standard Model particles, except for the case $p \rightarrow \pi^+ X$; this process requires at least one BSM fermion.}
 \label{Table:Symmetries}
\end{table}

For a simple model of this scenario, assume
a $B/L$-invariant theory in which one has added a neutral heavy scalar $\eta$ with $B=2$ and $L=2$; if this scalar gets a vacuum expectation value, then $U(1)_B \times U(1)_L$ is spontaneously
broken to $Z_4^{B+L} \times U(1)_{B-L}$. Note that the $U(1)_{B+L}$ of the underlying theory suffers from a quantum
anomaly, but this anomaly can be canceled by the addition of appropriate fermionic matter.

In an underlying higher energy theory, symmetries must permit processes in which both $B$ and $L$ increase or decrease by 2.  At quark level, these processes would involve 6 external quarks and 2 external leptons. The details of the Feynman diagrams would, however,
depend on the specific matter content and couplings of the theory. For example, if the matter content is appropriate, the
8 external fermions could couple in pairs to four scalars, which in turn couple to each other at a quartic vertex.
$B+L$ violation would be induced by a cubic scalar vertex of the form $\phi_i \phi_j \eta$, with the vev of $\eta$ inducing
mixing between scalars ($\phi_i \phi_j$) with the same $SU(3)_{QCD} \times U(1)_{EM}$ quantum numbers and $\Delta (B+L)= \pm 4$.

Some models with scalars inducing baryon and lepton number violation, without permitting protons to decay or annihilate to dileptons, were presented recently in Ref.~\cite{Arnold:2012sd}. Most work on signals of diproton to dilepton annihilation from grand unified and extended gauge theories was conducted over three decades ago \cite{Feinberg:1978sd,Mohapatra:1982aq,Mohapatra:1982aj,Arnellos:1982nt,Vergados:1982ef,Deo:1984tq,Axelrod:1985vy,Alberico:1986fd,Berger:1991fa}, with one recent exception considering an SU(2) extension of the Standard Model in which the charges of exotic heavy quarks are different than those of light Standard Model quarks \cite{Morrissey:2005uza}. Reference \cite{Morrissey:2005uza} includes hadron collider signatures of heavy-quark-exclusive diproton to dilepton annihilation processes, which thereby evade Frejus and Super-Kamiokande constraints, which are necessarily light-quark-exclusive. We will consider quark flavor specific $pp \to \ell^+ \ell^+$ in Section \ref{sec:hadroncolliders}.

\begin{figure}[h!]
\begin{center}
\begin{tabular}{cc}
\includegraphics[scale=.95]{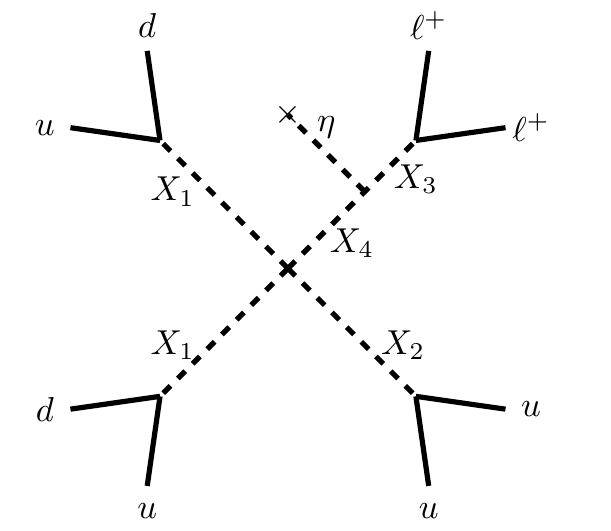}
&
\includegraphics[scale=.95]{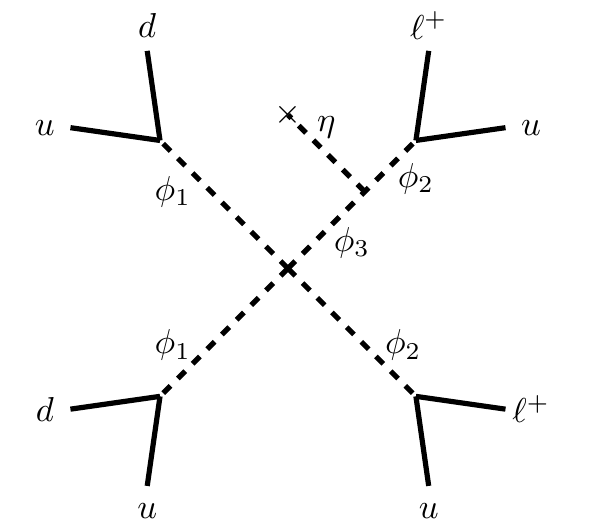}
\end{tabular}
\\
~~
\\
~~
\\
$(\rm{SU(3)_c , U(1)_{Y}}, U(1)_B, U(1)_L)$ charges
\\
\begin{tabular}{|c|c|}
\hline
\hline
\multicolumn{2}{|c|}{$\eta \in (1,0,2,2)$} \\
\hline
\hline
$X$ scalar fields & $\phi$ scalar fields \\
\hline
\hline
  $X_1 \in (3,-\frac{1}{3},-\frac{2}{3},0)$  & $\phi_1 \in (3,-\frac{1}{3},-\frac{2}{3},0)$ \\
 $X_2 \in (\bar{6},-\frac{4}{3},-\frac{2}{3},0)$ & $\phi_2 \in (\bar{3},\frac{1}{3},-\frac{1}{3},-1)$\\
  $X_3 \in (1,2,0,-2)$ &  $\phi_3 \in (\bar{3},\frac{1}{3},\frac{5}{3},1)$\\

 $X_4 \in (1,2,2,0)$ & \\
 \hline
\end{tabular}
\end{center}
\caption{Two interactions showing models of proton annihilation to dileptons along with the charge assignments of the scalar fields, $\eta$, $X_i$, and $\phi_j$. All charge assignments assume that quarks flow towards, while scalars and leptons flow away from the central quartic vertex.}
\label{fig:pptoll}
\end{figure}

Hereafter we present some $pp \to \ell^+ \ell^+$ models which break $U(1)_{B+L}$ down to a $Z_4$ symmetry with a single scalar, using quartic vertices of additional scalar fields with gauge charge assignments that couple diprotons to dileptons. These quartic vertices of scalars permitting $\Delta (B+L)= \pm 4$ are shown in Figure \ref{fig:pptoll}, along with a tabulation of the charges of the requisite scalars.
\renewcommand\arraystretch{1.3}

We will now examine how diprotons coupled to dileptons could be observed at experiments. The most interesting processes, from a phenomenological standpoint, are
\begin{itemize}
\item{$pp \rightarrow \ell^+ \ell^+$ -- observable as the isotropic deposition of $\sim 2~\gev$ of energy within Super-Kamiokande.  The related process $nn \rightarrow \bar \nu \bar \nu$ is constrained by observations of neutron stars. This is a {\it high-luminosity} signature.}
\item{$p e^- \rightarrow \bar p e^+$ -- hydrogen/anti-hydrogen mixing, observable as the isotropic deposition of $\sim 2~\gev$
of energy within Super-Kamiokande.  This process is also constrained by the formation of neutron stars.  This is a {\it high-luminosity} signature.}
\item{$qq \rightarrow \bar q \bar q \bar q \bar q \ell^+ \ell^+$ -- observable at the LHC as an excess of events with 4 jets, 2 like-sign
dileptons and little missing transverse momentum (the like-sign dileptons are predominantly positively charged).  This is a {\it high-energy} signature.}
\item{$pp \rightarrow \bar \phi \phi + X$ -- where $\phi$ is a QCD-charged heavy exotic particle which mediates $B$/$L$-violating
interactions.  This is a {\it high-energy} signature.}
\end{itemize}
Note that, there can be additional processes which respect the same symmetries, for example, $pp \rightarrow \ell^+ \ell^+ \pi^0 \pi^0$.  However,
as these processes would be expected to be subleading due to the corresponding phase space suppression, we can ignore them.

\subsection{Parameterizing the interactions with contact operators}

There is a wide range of freedom in constructing detailed models with an unbroken $Z_4^{B+L} \times U(1)_{B-L}$ global symmetry.
Instead, we will focus on parameterizing the effective interaction between Standard Model states and constraining the resulting
parameters with experiment and observation.  These parameterizations are relevant for processes at energies for which heavy exotic
particles cannot be produced.

At the low energies relevant for processes at Super-K, we can consider an effective theory where the degrees of freedom are protons, photons,
light mesons, and light charged leptons.  In this framework, we describe the processes
$p p \rightarrow \ell^+ \ell^+$  and $p e^- \rightarrow \bar p e^+$ in terms of contact operators made up of proton and lepton (assumed
to be an electron, for simplicity) bilinears,
\bea
\label{eq:Opp}
{\cal O} &=& \frac{1}{\Lambda^2} (\bar p^c \Gamma_p p) (\bar e^c \Gamma_e e),
\eea
where
\bea
\Gamma_{p,e} &=& 1, i \gamma^5, \gamma^\mu \gamma^5.
\eea
Thus, the possible operators are
\bea
{\cal O}_1 &=& {1 \over \Lambda_1^2} (\bar p^c  p) (\bar e^c  e)
\nonumber\\
{\cal O}_2 &=& {1 \over \Lambda_2^2} (i \bar p^c \gamma^5 p) (\bar e^c  e)
\nonumber\\
{\cal O}_3 &=& {1 \over \Lambda_3^2} (\bar p^c p) (i \bar e^c \gamma^5  e)
\nonumber\\
{\cal O}_4 &=& {1 \over \Lambda_4^2} (i \bar p^c \gamma^5 p) (i \bar e^c \gamma^5 e)
\nonumber\\
{\cal O}_5 &=& {1 \over \Lambda_5^2} (\bar p^c \gamma^\mu \gamma^5 p) (\bar e^c \gamma_\mu \gamma^5 e),
\label{eq:theops}
\eea
or any linear combination thereof.  The $\Lambda_i$ are parameters with units of energy, but it is
important to note that they should not be interpreted as the scale of the new physics which is integrated out
in the hadron-level effective field theory.  If there are new particles with mass $\gtrsim 1~\gev$, then the
hadron-level effective field theory would not be a good description of the coupling of the exotic particles with
Standard Model particles, and one would instead need a quark-level effective field theory.  Instead, the hadron-level
effective field theory is introduced simply as a convenient way of parameterizing bounds derived from low energy
experiments, for which the hadron-level effective field theory is valid; to relate this parametrization to a bound
on the scale of new physics, one should translate between the hadron-level and quark-level effective field theories.

For operators ${\cal O}_{1,3}$, $p p$ annihilation is
$p$-wave suppressed ($\propto v^2$), whereas for operator ${\cal O}_5$ it is chirality suppressed
($\propto m_e^2 / m_p^2$).
For a proton in a typical nucleus, $v \sim 0.1 c$, implying a $10^{-2}$ suppression of the nucleus decay rate if mediated exclusively
by a $p$-wave suppressed operator.
If the operator were chirality suppressed, the decay rate would be suppressed by $(m_\ell/ m_p)^2 \sim 10^{-2},10^{-6}$.  The constraint on the scale of the
$pp \ell^+ \ell^+$ contact operators derived from any bound on the nucleus decay rate are thus weakened by a factor of $\sim 3$ ($\Lambda_{1,3}$) and $\sim 30$ ($\Lambda_5$),
respectively.  As a benchmark we will assume that the relevant effective operator in the hadron-level effective field theory is ${\cal O}_4$, which is
$CP$-invariant and permits unsuppressed proton self-annihilation; we will show that deviations from this assumption will not significantly affect bounds on the scale of new physics.

For collider processes, the relevant Standard Model degrees of freedom are quarks and leptons.  If the
mass of the mediating particles are much larger than the collider energy scale, the relevant interaction between
six quarks and two leptons can be expressed in terms of a dimension 12 contact operator.  Examples of such operators
include
\bea
{\cal O}_{Q1} &=& {1 \over \Lambda_{Q1}^8} (\bar Q^c P_L Q)^2 (\bar Q^c P_L l) (\bar \ell^c P_L Q),
\label{eq:Oq1}\\
{\cal O}_{Q2} &=& {1 \over \Lambda_{Q2}^8} (\bar Q^c P_L Q)^2 (\bar u_R^c P_R e_R) (\bar e_R^c P_R u_R),
\\
&~&\rm{etc.},
\nonumber
\eea
where the 6 quark fields are implicitly color-contracted by two $\epsilon^{abc}$ tensors.
The coefficient $\Lambda_{Qi}$ can be bounded with collider experiments, and is roughly related
to the scale of new physics.

To compare collider constraints on the $\Lambda_Q$ with constraints on the $\Lambda$ arising from low-energy data,
one must match the coefficients of the operators which arise in the two different descriptions.
One can do this, for example, by equating the matrix element for the process $p e^- \rightarrow \bar p e^+ $,
as computed in each description.
The two expressions for the matrix element for $p e^- \rightarrow \bar p e^+ $ (in the non-relativistic limit) are given by
\bea
{\cal M}_{Qi}(p e^- \rightarrow \bar p e^+ ) &\propto & {m_e m_p^7\over \Lambda_{Qi}^8}
\nonumber\\
{\cal M}_{i}(p e^- \rightarrow \bar p e^+ ) &\propto & {m_e m_p\over \Lambda_{i}^2}
\label{eq:twops}
\eea
The quark operators are
acting on protons states, but in those terms $m_p$ is the only relevant scale (in the limit where the
$u$- and $d$-quarks are massless), so the dependence on $m_p$ is just determined by dimensional analysis.

To understand more precisely the dependence on $m_p$,
we can use the following measurements of the hadronic parameter $\beta_H$, which in
our relativistic state normalization can be written as
\bea
\langle 0 | u_L u_L d_L |p \rangle &=& \sqrt{2 m_p} \beta_H
\eea
where $\beta_H = 0.003~\gev^3$~\cite{Donoghue:1982jm} or
as large as $\beta_H = 0.014~\gev^3$~\cite{Kuramashi:2000hw}.  This matrix
element is precisely relevant for ${\cal O}_{Q1}$, so for our purposes is a reasonable choice to account for
QCD effects.\footnote{The same result is obtained by replacing $m_p$ with the QCD scale, $m_p \rightarrow \Lambda_{QCD} = 0.22~ \rm{GeV}$.}
We see that adopting either of these values would
result in rescaling $m_p$ in our decay rate by $0.22$ or $0.34$, respectively.

Equating the matrix elements in Eq.~\ref{eq:twops}, and rescaling $m_p$, we find the correspondence
\bea
\Lambda (0.22 m_p)^{3} &\simeq & \Lambda_{Qi}^4 .
\label{eq:matching}
\eea
This correspondence can also be derived simply from the fact that the effective quark-level operator
is dimension 12, while the effective hadron-level operator is dimension 6.  We see that even a factor
$\sim 30$ change in the bound on the hadron-level parameter $\Lambda$ arising from low energy experiments
results in only a factor $\sim 2$ change in the bound on the scale of new physics, justifying our
simplifying choice of ${\cal O}_4$ as the only contact operator in the hadron-level theory.

\subsection{Energy versus luminosity}

The above relation leads to an interesting
correlation between the energy, luminosity and sensitivity.  For any type of experiment, the number of
events mediated by this effective operator scales as
\bea
N \propto \frac{{\cal L} }{ E^2} \left(\frac{E }{ \Lambda_Q} \right)^{16} \label{eq:relation}
\eea
where $\Lambda_Q$ is the scale of new physics, $E$ is the energy scale of the process, and
${\cal L}$ is the integrated luminosity of the experiment.  For a low energy process, such as proton annihilation
in Super-Kamiokande, $E \sim m_p$, and the integrated luminosity ${\cal L}$ is related to the
density of the material, the fiducial volume, and the period of time studied.  For a high-energy collider process, however, ${\cal L}$ is the
related to the luminosity of the beams, while $E$ is determined largely by the collider energy.  This scaling
relation holds for any process in which the heavy mediating particles are not directly produced, and the only final
state particles are Standard Model particles.

For low-energy experiments, the large suppression resulting
from the low energy of the process must be compensated by the very large effective exposure.  It may well be the
case that these experiments provide the greatest current sensitivity to new physics, but one immediately sees that
the sensitivity of these experiments cannot increase very much; $E$ is fixed, and the sensitivity to $\Lambda_Q$ scales
only as ${\cal L}^{1/16}$.  By contrast, one sees that collider experiments cannot increase their sensitivity very much
with higher luminosity, but can with higher energy.  Note also that, for a given number of events, the scale of new
physics is largely insensitive to changes in the overall proportionality coefficient.  As a result, suppressions
arising from details of the particle physics model, such as $p$-wave or chirality suppression, will have negligible
impact on the sensitivity of any given experiment.

We may compare the above case to the
case where $\Delta L =0$, $\Delta B=\pm 2$, allowing the process $pp \rightarrow \pi^+ \pi^+, K^+ K^+$.  For these
processes, the quark-level operator is dimension 9.  As a result, one would find that the number of events at an experiment scales as $N \propto ({\cal L}/E^2)(E /\Lambda_Q)^{10}$.  Again we see for $pp \rightarrow \pi^+ \pi^+, K^+ K^+$ that an increase in luminosity does not improve sensitivity significantly, while an increase in the energy of the process does.

On the other hand, the actual sensitivity of rare event searches to the scale of new physics does depend strongly on the specific process.
Rare event (water Cherenkov) searches for proton annihilation to dileptons and dimesons search for an excess of events above the
estimated background for the exposure of the experiment. If an experiment yields no statistically significant excess of events, the number of allowed events for both $pp \rightarrow \pi^+ \pi^+$ and $pp \rightarrow \ell^+ \ell^+$ processes is about the same, because rare event searches are not very sensitive to the topology of the final state, provided it is isotropic and has a large energy deposition. Thus, because the number of $pp \rightarrow \pi^+ \pi^+$ events scales with a smaller power of $(E /\Lambda_Q)$, rare event searches like those conducted at Super-K will constrain the scale of new physics for the process $pp \rightarrow \pi^+ \pi^+$ more tightly than for $pp \rightarrow \ell^+ \ell^+$.

\section{Bounds from Super-Kamiokande}
\label{sec:superk}

Super-Kamiokande can bound the effective quark-lepton interaction operators in several ways.
First, the process $pp \rightarrow \ell^+ \ell^+$ ($l =e, \mu$) can allow the decay of an oxygen nucleus.  Secondly,
the process $p e^- \rightarrow \bar p e^+$ can allow either hydrogen annihilation or the annihilation
of a proton in an oxygen nucleus against an inner shell electron.

Both of these processes yield a striking signature. The process $p e^- \rightarrow \bar p e^+$ causes an antiproton to rapidly annihilate with a proton and deposit $2~\gev$ of energy isotropically
within the detector.  This is also the signature used in Super-Kamiokande's analysis of  $n - \bar n$
oscillation~\cite{Abe:2011ky}, which found $24$ candidate events and yielded a bound on the neutron lifetime of
\bea
T_{n - \bar n} &>& 1.89 \times 10^{32}~\yr.
\eea
The estimated number of background events satisfying the selected cuts was $24.1$.

A Super-Kamiokande study of $pp \to \ell^+ \ell^+$ would result in two ultra-relativistic, back-to-back, same-sign leptons with clearly identifiable Cherenkov cones. In the case of proton annihilation to antimuons, two widely separated $\mathcal{O}(5)$ meter muon tracks with a common vertex is likely to have a small background. These two muons coming off back to back and ending in muon decays (since the  $\mu^+$ is not rapidly absorbed), with the muons each forming long tracks will be particularly distinct from $\nu_\mu$ charged current interactions; $\sim$GeV energy atmospheric neutrinos will be strongly forward, and rarely will a backwards traveling muon be generated. The bound on lifetime is likely to be comparable to or stronger than the recent limit, $T_{p \to K^+ \nu} > 6 \times 10^{33}~\yr$ which was set for $p \to K^+ \nu$ \cite{Abe:2014mwa}. However, to remain conservative until a Super-Kamiokande $pp \to \ell^+ \ell^+$ study is completed, we will use the Super-Kamiokande bound on $n - \bar n$ oscillations to set our bound on proton annihilation to dileptons.\footnote{Using a fiducial volume of iron interleaving Geiger tubes, the Frejus collaboration \cite{Berger:1991fa} set a direct bound on proton annihilation to dileptons, $T_{pp \to \ell^+ \ell^+} \gtrsim 10^{31}~\yr$. However, this is less stringent than the bound inferred from Super-K limits on neutron oscillation.}

Of course, the signatures of the processes we consider may differ from the expectations outlined here; for example, the process $p e^- \rightarrow \bar p e^+$ yields a low-energy positron, which we assume is undetectable in the midst of a 2 GeV hadronic spray of the annihilating proton-antiproton pair. To properly establish the signature and background, Monte Carlo simulations would need to be conducted. However, the details of the cuts and signature are relatively unimportant for our purposes here, as the sensitivity to new physics depends only weakly on the number of events required for statistical significance.  As such we are justified in simply taking the Super-Kamiokande bound on the number of events from $\Delta (B-L)=0$, $\Delta (B+L)= \pm 4$ processes to be $\lesssim {\cal O}(10)$, occurring at a rate $\Gamma \lesssim {\cal O}(10^{-31})~\yr^{-1}$.   The corresponding bounds on the lifetime of oxygen and hydrogen are $T_O \gtrsim 2.4 \times 10^{31}~\yr$ and $T_H \gtrsim 4.7 \times 10^{31}~\yr$, respectively.

\subsection{$pp \rightarrow \ell^+ \ell^+$ }

The only $CP$-invariant contact operator which permits $pp$ annihilation that is neither $s$-wave nor
chirality suppressed is ${\cal O}_4$, and we will use this operator as a benchmark.
The amplitude for non-relativistic proton-proton annihilation via ${\cal O}_4$ is
\bea
{1 \over 4 } \sum_{spins} |{\cal M}_{p p \rightarrow \ell^+ \ell^+}^{{\cal O}_4}|^2
\sim 16\left({m_p^4  \over \Lambda_4^4} \right).
\eea
If we crudely approximate the protons of a ${}^{16}O$ nucleus to have a uniform spherical distribution cut off
at radius $r = 3~\fm$, then the rate for the process ${}^{16}O \rightarrow {}^{14}C +\ell^+ \ell^+$ is
\bea
\Gamma_{{}^{16}O \rightarrow {}^{14}C \ell^+ \ell^+}^{{\cal O}_4}
&=&
{21  \over 4 \pi^2 }  \left({m_p^2  \over \Lambda_4^4} \right)   r^{-3}
\nonumber\\
& = & (6.4 \times 10^{27}~\yr^{-1}) \left(\frac{\gev}{\Lambda_4} \right)^{4}.
\eea
We can use this estimate of the annihilation rate for protons in oxygen nuclei to bound the coefficient of ${\cal O}_4$ with Super-K neutron oscillation constraints. This yields the constraint $\Lambda_4 > 7 \times 10^{14} ~\gev$.
The corresponding bound on the scale of quark-level contact operator
is $\Lambda_Q > 1.6~\tev$ (eq.~\ref{eq:matching}), with some uncertainty due to the hadronic matrix element as discussed in Section \ref{sec:loweproc}.

\subsection{$p e^- \rightarrow \bar p e^+$}
\label{sec:hydann}

The operators given in \eqref{eq:theops} also allow hydrogen to annihilate via the process $p e^- \rightarrow \bar p e^+$.
For simplicity we consider the operator ${\cal O}_1$,
which can be rewritten (using a  Fierz transformation) as
\bea
{\cal O}_1 &=&  {1 \over 4\Lambda_1^2} \left[ (\bar e^c  p) (\bar p^c  e)
-(i \bar e^c \gamma^5 p) (i \bar p^c \gamma^5 e)
+(\bar e^c \gamma^\mu p) (\bar p^c \gamma_\mu e)
\right.
\nonumber\\
&\,& \left.
-(\bar e^c \gamma^\mu \gamma^5 p) (\bar p^c \gamma_\mu \gamma^5 e)
+(\bar e^c \sigma^{\mu \nu} p) (\bar p^c \sigma_{\mu \nu} e) \right].
\eea
Because the Fierz transformation yields a sum of operators with many Lorentz structures and no cancelations, we see that
similar bounds on the process $p e^- \rightarrow \bar p e^+$ would result from any other contact operator
${\cal O}_i$.

From Appendix B of~\cite{Kumar:2013iva}, for example, we see that the only terms relevant for the annihilation of
an $S=0$, $L=0$ initial state (which composes the vast majority of naturally occurring hydrogen) are
$(i \bar e^c \gamma^5 p) (i \bar p^c \gamma^5 e)$ and $(\bar e^c \gamma^\mu \gamma^5 p) (\bar p^c \gamma_\mu \gamma^5 e)$.
The decay rate for hydrogen induced by this operator is given by
\bea
\Gamma_1^H &=&
{\alpha^4 m_e^5 \over 2 \pi^2 \Lambda_1^4} = (7.6 \times 10^{-15}~\s^{-1})
\left({\Lambda_1 \over \tev} \right)^{-4} .
\eea

A similar calculation yields the rate for oxygen decay through annihilation of a $1S$ electron
against a proton in the nucleus.  For simplicity, we can estimate the decay rate of oxygen by treating oxygen as a
hydrogen-like system with two $1S$ electrons.  We thus assume that the
outer electrons contribute negligibly to oxygen decay, and that the wavefunction of the $1S$ states are not
significantly affected by the other oxygen electrons. We then find
\bea
\Gamma_1^O &\sim&
{(Z\alpha)^4 (16) m_e^5 \over 2 \pi^2 \Lambda_1^4} \sim 2^{16} \Gamma_1^H
\eea

From the limit on the hydrogen lifetime derived from Super-Kamiokande's $n - \bar n$ oscillation
data, we conclude
\bea
\Lambda_1 &\geq&  1.8 \times 10^9~\gev,
\eea
and when we include the decay rate of oxygen, this bound on $\Lambda_1$ increases by a
factor of $\sim 2^4 =16$. Regardless, we find that constraints arising from $p e^- \rightarrow \bar p e^+$
are generically weaker than those arising from the process $pp \rightarrow \ell^+ \ell^+$ occurring
within an oxygen nucleus. In particular, even if the effective contact operator yields a
proton annihilation matrix element which is $p$-wave or chirality suppressed, the bound on
the energy scale $\Lambda$ from $pp$-annihilation is still more stringent that that arising from
$pe^-$-annihilation, even accounting for uncertainties in the nuclear wavefunction.

\section{Bounds on dinucleon-dilepton coupling from neutron stars}
\label{sec:nstars}

Neutron stars can yield two types of bounds on $\Delta (B-L)=0$, $\Delta (B+L)= \pm 4$ processes.  First, as the neutron
star forms, the protons and electrons become more dense
and the rate for the process $p e^- \rightarrow \bar p e^+$ increases, which could disrupt neutron
star formation.  Secondly, once the neutron star has formed, the process $n n \rightarrow \bar \nu \bar \nu$
may be allowed, and would be related to $p p \rightarrow \ell^+ \ell^+$ at quark-level by weak isospin.

\subsection{Neutron star bounds on dineutron annihilation}

The quark-level effective operators which permit the process $p e^- \rightarrow \bar p e^+$, if they
involve the factor $(\bar Q^c P_L l_L)$, also permit $nn$-annihilation.
The largest annihilation cross section would arise if $nn \rightarrow \bar \nu \bar \nu$ proceeded 
through a pseudoscalar interaction at hadron level, such as ${\cal O}_4$ after the 
replacement $p \rightarrow n$, $e \rightarrow \nu$.  This scenario involves coupling to neutrinos 
of both helicities, and would thus be viable in the case of
Dirac neutrinos. 
The rate for a neutron to annihilate can be then estimated by

\bea
\Gamma_{nn}^{NS} &\sim&  {\eta m_p^2 \over 8\pi    } \left({1  \over \Lambda_4} \right)^4
\nonumber\\
&\sim& (10^{20}~\s^{-1}) \left(\Lambda_4 \over \gev \right)^{-4}
\eea
where $\eta \sim 2 \times 10^{53} / \rm{km}^3 \approx 0.002~\gev^3$ is the neutron number density in a 1.4 solar mass
neutron star. If we have $\Lambda_4 \geq 10^{10}~\gev$, then we find $\Gamma_{nn}^{NS} \leq 10^{-20}~\s^{-1}$, implying that
$nn$-annihilation would have very little effect on the star, even over the age of the universe.

\subsection{Neutron star formation bounds on hydrogen annihilation}

The rate for a single proton to annihilate via the process $p e^- \rightarrow \bar p e^+$ during neutron
star formation is similar to the rate for a single neutron to annihilate after the neutron star has
formed, because the relevant number densities and energy scales of the two processes are roughly similar.
But the electron density of a forming neutron star is depleted over a time scale $\sim {\cal O}(10)~{\rm s}$~\cite{Janka:2000ce,Janka:2004np},
implying that this process has negligible impact on the formation of the neutron star.

We thus see that observations of neutron stars do no place competitive bounds on the $B/L$-violating
couplings which we consider, owing to the large number of events required to yield an observable effect
in an astrophysical body. The preceding treatment has assumed a neutron star of constant density. However, a more in-depth study might consider the possibility that rapid high-energy processes such as pulsar glitches \cite{Espinoza:2011pq} or strong force somnoluminescence in neutron stars \cite{Simmons:1996ks} could place a tighter bound on proton annihilation to dileptons.

\section{Diproton - dilepton coupling at the LHC and beyond}
\label{sec:hadroncolliders}

In this section we determine bounds on $\Lambda_{Q1}$ from colliders.  If
the scale of new physics is large compared to the characteristic energy scale
of hard processes at the collider, then the contact approximation is valid and
we may parameterize the $B/L$-violating interactions via the dimension 12 quark level
effective contact operators which we have described.
Indeed, for simple models with only a few
mediating particles, one would expect $\Lambda_Q$ to be related to the mass scale $M$ of new particles by the relation
$M \sim g \Lambda_Q $, where $g$ is the coupling of the new particles.
At colliders, the operator ${\cal O}_{Q1}$ will allow processes like
$pp \rightarrow \ell^+ \ell^+ +{4\,\rm{jets}}$ at the LHC, or $e^- p \rightarrow e^+ +{5\, \rm{jets}}$ at HERA.
Note that if we instead consider the other operators ${\cal O}_{Qi}$, one might expect event rates which could
differ by ${\cal O}(1)$ factors; but given that $N \propto \Lambda_Q^{-16}$, our choice of ${\cal O}_{Q1}$ does not significantly alter LHC sensitivity.

However, if the energy scale of hard collider processes is larger than the scale of new physics, the contact
approximation will fail.  Instead, the mostly promising search strategy will be via production of
the exotic mediators.  In simple models, such as the ones we described, at least some
of the mediating particles are charged under $SU(3)_{QCD}$, and thus are easily produced at hadron colliders
provided that the center of mass energy is sufficient to pair produce the exotic particles.
Direct searches at the LHC currently constrain this mass scale to
satisfy $M \gtrsim {\cal O}(500)~\gev$ \cite{Chatrchyan:2014lfa,Aad:2014bva}.

\subsection{$pp \rightarrow \ell^+ \ell^+ +$4 jets}

We first determine the sensitivity of the LHC to this operator by numerical simulation of the process $pp \rightarrow \ell^\pm \ell^\pm
+{4\,jets}$.\footnote{One might also consider dimension twelve processes related by weak isospin like $\bar{u}\bar{u} \to dddd \nu_e \nu_e $,
yielding $\geq 4 ~{\rm jets}$ + missing transverse energy (MET). The cross-section for these will be the same as for the process listed 
in eq.~\eqref{eq:uulhc}, but with
much larger hadronic backgrounds, making this final state less incisive.}
The relevant parton-level hard processes are:
\begin{eqnarray}
u ~u &\to & \bar{u}~ \bar{u}~ \bar{d}~ \bar{d}~ e^+~ e^+ ~+ {\rm h.c.}, \nonumber \\
u ~d &\to & \bar{u}~ \bar{u}~ \bar{u}~ \bar{d}~ e^+~ e^+ ~ + {\rm h.c.},\nonumber \\
{\rm and} ~~d ~d &\to & \bar{u}~ \bar{u}~ \bar{u}~ \bar{u}~ e^+~ e^+ ~+ {\rm h.c.},
\label{eq:uulhc}
\end{eqnarray}
at the LHC.
The major Standard Model backgrounds for the $\ell^\pm \ell^\pm + 4 ~{\rm jets}$ channel are $t \bar{t} W$ production where the $W$ and a top quark decay leptonically, along with $W + {\rm jets}$ or $t \bar{t}(j)$ with the $W$ or top decaying leptonically, and a jet faking a lepton. The backgrounds with fake leptons are often identified in experimental studies as ``non-prompt" lepton backgrounds \cite{Chatrchyan:2013fea,ATLAS-CONF-2012-105,Aad:2014pda,Khachatryan:2014qaa}.

In Table \ref{tab:ssl4j}, we show the number of signal events in the $\ell^\pm ~\ell^\pm  + 4 ~{\rm jets}$ channel \cite{Chatrchyan:2013fea} arising from the
dimension-twelve operator ${\cal O}_{Q1}$ given in Eq.~\eqref{eq:Oq1}, assuming
either first or second generation quarks or leptons.  We also list the relevant detector cuts, the number of observed events, and the number of expected
background events.
The exact signal region of this study that is most constraining for a dim-12 $\Delta(B+L)=\pm 4$ operator is the ``High-$p_T$" lepton pre-selected ``SR04" region, a region which requires two same-sign leptons with transverse momentum greater than 20 GeV, and four non $b$-jets with a $p_T$ sum greater than $400 ~{\rm GeV}$, along with other cuts indicated in Table \ref{tab:ssl4j}.
This data yields the bound
$\Lambda_{Q1} \geq 830 ~ {\rm GeV}$ with $\sim 2 \sigma$ significance.

It is important to note that the details of the cut choice and background model are relatively unimportant because the analysis
is essentially signal limited.  As seen in Table \ref{tab:ssl4j}, and as may be expected from any signal with the topology of like sign di-leptons with
no missing energy, the number of background events is small.  Given that the number of signal events scales as $N \propto \Lambda_Q^{-16}$, even a change of
a few orders of magnitude in the number of events required for exclusion would only change the collider sensitivity reach by an ${\cal O}(1)$ factor.  Thus,
for an analysis of either the LHC or future high-energy colliders, it is sufficient to ignore any detailed modeling of cuts, backgrounds or detector
performance, and instead simply require a few signal events with parton level cuts. Complete details of the collider analysis conducted here can be found in Appendix \ref{app:collider}.

\begin{table}[h!]
\begin{center}
\begin{tabular}{c}
\begin{tabular}{|l|| r| r| r| r| r| r| r|}
\hline
Cutoff $\Lambda_{Q1}$ in GeV for $\sqrt{s}=8~\rm{TeV}$, $20~{\rm{fb}^{-1}}$ & 800 & 830 & 870 & 900 & 940 & 980 & 1020 \\
\hline
\hline
Events from (u,d,e) in Eq. \eqref{eq:Oq1} & 12 & 6 & 3 & 1.6 & 0.8 & 0.45 & 0.25 \\
\hline
Events from (u,d,e,${\rm \mu}$) in Eq. \eqref{eq:Oq1} & 30 & 15 & 7.5 & 4 & 2 & 1 & 0.6 \\
\hline
\end{tabular}
\\
\\
\begin{tabular}{|l||  r| r| r| r| r| r| r|}
\hline
Cutoff $\Lambda_{Q1}$ in GeV for $\sqrt{s}=8~\rm{TeV}$, $20~{\rm{fb}^{-1}}$ & 440 & 470 & 510 & 550 & 580 & 620 & 650 \\
\hline
\hline
Events from (c,s,e) in Eq. \eqref{eq:Oq1} & 54 & 15 & 4.5 & 1.5 & 0.5 & 0.2 & 0.1 \\
\hline
Events from (c,s,e,${\rm \mu}$) in Eq. \eqref{eq:Oq1} & 103 & 31 & 9 & 3 & 1 & 0.4 & 0.15 \\
\hline
\end{tabular}
\\
\\
\begin{tabular}{|c|}
\hline
Result for signal region ``SR04" in Ref.~\cite{Chatrchyan:2013fea}\\
\hline
\hline
2 measured events, 5.6 $\pm$ 2.1 ($1 \sigma$) expected \\
\hline
\end{tabular}
\\
\\
\begin{tabular}{|c|}
\hline
Cuts for signal region ``SR04" of Ref.~\cite{Chatrchyan:2013fea} and this study \\
\hline
\hline
4 jets (light quarks) with $p_T > 40 ~{\rm GeV}$ \\
50 GeV $<$ MET $<$ 120 GeV \\
2 same-sign leptons with $p_T > 20 ~{\rm GeV}$ \\
$H_T \equiv \sum_{\rm{jets}} |p_T| > 400 ~{\rm GeV}$\\
\hline
\end{tabular}
\end{tabular}
\end{center}
\caption{These tables give the number of events expected in the $\sqrt{s}=8 ~\rm{TeV}$, 20 $\rm{fb}^{-1}$, LHC data from the dimension twelve $B+L$ violating operator of Eq.~\eqref{eq:Oq1} for $\mathcal{O}(1)$ couplings and a cutoff $\Lambda_{Q1}$ in the $\ell^\pm ~\ell^\pm  + 4 ~{\rm jets}$ detection channel. The expected events are given for either six first generation quarks (u,d) or six second generation quarks (c,s) coupled to either two same-sign electrons only or both electrons and muons.}
\label{tab:ssl4j}
\end{table}

Although Super-Kamiokande studies of higher-dimension baryon and lepton number violating operators are necessarily restricted to first generation quarks,
collider studies have a broader scope. The results in Table \ref{tab:ssl4j} show the bound and reach of the 8 TeV LHC for the $B+L$ violating operator of Eq.~\eqref{eq:Oq1} composed of either first ($u$,$d$) or second ($c$,$s$) generation quarks coupled to either electrons only or both electrons and muons. We note that while operators composed of first-generation quarks are more tightly constrained by Super-Kamiokande than by the LHC, operators coupling to second generation quarks which violate $B+L$ mod four will be entirely unconstrained by Super-Kamiokande. This motivates models of $B+L$ violation exclusive to second and third generation quarks: such models could be discovered at future runs of the LHC, whereas similar first generation quark operators may
require a more energetic proton collider, given Super-Kamiokande bounds. For other research which includes flavor-specific baryon and lepton number violation prospects at the LHC, see \cite{Baldes:2011mh}.

\subsection{Comparison to searches for direct production of exotic colored mediators}

We have seen that the bound set by the LHC on contact operators is already in a regime where the contact approximation is expected to begin breaking down. As a result, the LHC bound should be treated as somewhat heuristic, and really indicative of the fact that the LHC has not accumulated enough integrated luminosity to probe new physics for which the contact approximation would be clearly
appropriate (unless the new physics involved very large numbers of mediators).
In fact, because the number of signal events
at the LHC scales as $N \sim ({\cal L} /E^2)(E / \Lambda_Q)^{16}$, it will never be practically possible to obtain
enough luminosity for the contact approximation to be valid. It is thus useful to consider the contrast between a
high-energy search focused on the production of new particles, as opposed to a high-luminosity search for the indirect
effects of intermediate heavy mediators. We see that for the scenario considered here, only a high-energy search strategy
is practical, and the key targets are heavy exotic $SU(3)_{QCD}$-coupled particles.

\begin{table}[t!]
\begin{tabular}{c}
\begin{tabular}{|l||  r| r| r| r| r| r|}
\hline
Cutoff $\Lambda_{Q1}$ in GeV for $\sqrt{s}=14~\rm{TeV}$, $3~{\rm{ab}^{-1}}$ & 1300 & 1450 & 1600 & 1750 & 1900 & 2050 \\
\hline
\hline
Events from (u,d,e,${\rm \mu}$) in Eq. \eqref{eq:Oq1} & 3652 & 676 & 147 & 36 & 10 & 3.1 \\
\hline
\end{tabular}
\\
\\
\begin{tabular}{|l|| r| r| r| r| r| }
\hline
Cutoff $\Lambda_{Q1}$ in TeV for $\sqrt{s}=100~\rm{TeV}$, $10~{\rm{ab}^{-1}}$ & 7 & 9 & 10 & 11 & 13 \\
\hline
\hline
Events from (u,d,e,${\rm \mu}$) in Eq. \eqref{eq:Oq1} & 8511 & 459 & 39 & 4.6 & 0.7  \\
\hline
\end{tabular}
\end{tabular}
\label{fig:LHCand100}
\caption{These tables show the expected number of dinucleon-dilepton coupled events from operator \eqref{eq:Oq1} at both a high-luminosity run of the LHC and a high-luminosity run of a future 100 TeV collider. This result assumes the collider acceptance of partons found in an 8 TeV LHC study \cite{Chatrchyan:2013fea} of $\ell^\pm \ell^\pm 4j$ events.}
\label{tab:futurecolliders}
\end{table}

We now consider how detection prospects for dinucleon-dilepton coupling will improve at a high-luminosity run of the LHC, and at a future $\sqrt{s}=100$ TeV collider. To estimate the sensitivity of the these machines, we passed parton-level events through the same lepton, jet, and MET acceptances reported in \cite{Chatrchyan:2013fea}. A precise prediction of constraints on \eqref{eq:Oq1} would require a detailed examination of relevant hadronic backgrounds to $\ell^\pm \ell^\pm + 4j$ (including lepton fake rates), and specifically how these backgrounds scale with increased energy and the particulars of detectors at a very high-luminosity LHC and a future 100 TeV collider.
However, even a naive extrapolation of backgrounds and systematic uncertainties (described in the appendix) is sufficient for our purposes, as the sensitivity
is only weakly dependent on the number of events needed for detection, which we can take to be ${\cal O}(1)$.

In Table \ref{tab:futurecolliders} we show the number of signal events (similar to Table \ref{tab:ssl4j}) for either a
$14~\tev$ hadron collider with $3 \ab^{-1}$ integrated luminosity, or a $100~\tev$ hadron collider with $10~\ab^{-1}$ integrated
luminosity.
A 100 TeV collider could access dinucleon-dilepton coupling processes mediated by new physics at an energy scale of 10 TeV,
putting its sensitivity beyond that of both Super-Kamiokande and the projected future reach of
Hyper-Kamiokande~\cite{Abe:2011ts}.\footnote{The projected order of magnitude longer lifetime constraint on dinucleon decay processes in \cite{Abe:2011ts}
implies an $\mathcal{O}(1)$ shift in the allowed quark-level cutoff of (\ref{eq:Oq1}).}
Again, we note that if these interactions are mediated by
new physics appearing at the energy scale $\sim 10~\tev$, the contact approximation would not be expected to be valid at a $100~\tev$ hadron collider.
Instead, this result is indicative of the fact that the one should instead expect the promising search strategy to be the production of heavy exotic colored particles,
to which a $100~\tev$ collider should be sensitive provided their mass is $\lesssim 20~\tev$ \cite{Yu:2013wta,Anderson:2013ida,Fowlie:2014awa,Low:2014cba,Cohen:2013xda,Larkoski:2014bia,Cohen:2014hxa,Cirelli:2014dsa,Curtin:2014jma,Acharya:2014pua,Alves:2014cda,Gori:2014oua}.

\begin{figure}[t!]
\begin{center}
\includegraphics[scale=1.2]{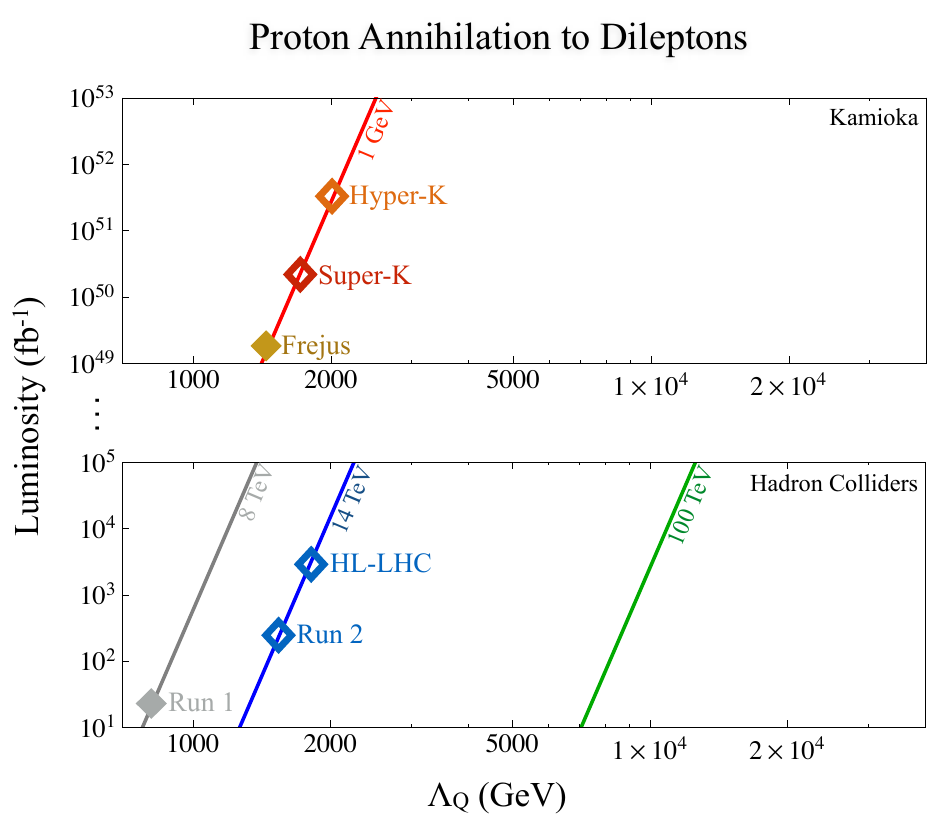}
\end{center}
\caption{A plot of hadron collider and Kamioka discovery prospects for $pp \to \ell^+ \ell^+$, as determined by sensitivity to the cutoff of $(qqq\ell)^2 / \Lambda_Q^8$ (see Eq. \ref{eq:Oq1}). The projections were arrived at using Eq. \ref{eq:relation} and the results of Sections \ref{sec:superk} and \ref{sec:hadroncolliders}.}
\label{fig:energyvslum}
\end{figure}

This highlights the unique prospects available to higher energy hadron colliders. $\mathcal{O}(100~\tev)$ machines could surpass water Cherenkov detection of rare baryon and lepton number violating processes, but this is \emph{only} attainable through a sizable increase in center-of-mass energy, as illustrated in Figure \ref{fig:energyvslum}. Furthermore, in a scenario with $\Delta B= \pm 2$, $\Delta L=0$ (allowing $pp \rightarrow \pi^+ \pi^+$), even higher energies would be necessary for colliders to be competitive. For this scenario, current bounds from Super-Kamiokande constrain the scale of new physics at the level $\Lambda_Q \gtrsim m_p (1.5~\tev / m_p)^{{16 \over 10}} \sim 10^5~\gev$.  One would need a PeV-scale hadron collider (a Pevatron) to obtain bounds on
the production of the heavy mediating particles which are competitive with the sensitivity of rare event searches.
The process $pp \rightarrow e^+ e^+, \mu^+ \mu^+$ thus provides unique opportunities for foreseeable high-energy hadron collider experiments.

It is worth noting that cosmic ray protons, whose energies are constrained by the GZK cutoff, can annihilate against protons in the atmosphere at center-of-mass energies
as large as $10^6~\gev$.  Although this may be the highest energy environment for probing processes with $\Delta (B+L) =\pm 4$, there seems no practical way of
distinguishing such events from much more common and prosaic events, such as $p \bar p$-annihilation.  As such, cosmic ray studies are not likely to be a
fruitful method for probing such $B/L$-violating processes.

\section{Conclusion}
\label{sec:conclusion}
\begin{center}
\begin{table}[t!]
\begin{tabular}{|c|c|c|c|}
  \hline
  experiment & number of events & process energy & effective luminosity \\
  \hline
  Super-K & ${\cal O}(10)$ & $\sim 1~\gev$ & $\sim 10^{50}~\fb^{-1}$ \\
  neutron star &  $\sim 10^{55}$ & $\sim 1~\gev$ & $\sim 10^{83}~\fb^{-1}$ \\
  LHC & ${\cal O}(1)$ & $\sim 10^3~\gev$ & $\sim 10^2~\fb^{-1}$ \\
  HE-LHC & ${\cal O}(1)$ & $\sim 10^4~\gev$ & $\sim 10^3~\fb^{-1}$ \\
  \hline
\end{tabular}
\caption{A description of various experiments, along with the
energy scale of the relevant process, the effective luminosity, and the
number of events needed for statistical significance.  For Super-K and neutron
star observations, the effective integrated luminosity is given by ${\cal L} \sim (1/2) N \eta v t$,
where $N$ is the number of nucleons in the target, $\eta$ is the local nucleon number
density, $t$ is the time the nucleons are observed, and $v$ is their average velocity.}
\label{tab:scaling}
\end{table}
\end{center}
In this work, we have considered a general class of scenarios in which $B-L$ is conserved, but $B+L$ is only
conserved modulo 4. This symmetry structure guarantees the absolute stability of the proton, but permits
proton annihilation via the process $pp \rightarrow \ell^+ \ell^+$. We have found that these processes can be
constrained by searches for rare processes at Super-Kamiokande, by observations of neutron stars, and by
LHC searches for like-sign dileptons accompanied by many jets.

Since processes with $\Delta (B-L)=0$, $\Delta (B+L)=\pm 4$ involving only Standard Model particles are mediated
by a dimension-12 quark-level operator, we find that the number of such events scales as
$N \propto ({\cal L}/E^2)(E/\Lambda)^{16}$, where $E$ is the energy scale of the process, ${\cal L}$ is effective
integrated luminosity, and $\Lambda$ is the scale of new physics.  In Table~\ref{tab:scaling} we list approximate values of $N$, $E$,
and ${\cal L}$ for various experimental environments.  We see that the extremely large effective luminosity of
experiments such as Super-Kamiokande allow it to have the greatest sensitivity to new physics, despite the modest energy
of the relevant process.  Although neutron stars have a much greater effective luminosity, the extremely large
number of events needed to probe new physics (we assume that the annihilation of 1\% of the nucleons is required)
weakens the sensitivity of this strategy; the quantity ${\cal L} / N$ will always be larger for an earth-based rare
event search than for an astronomical observation.
An $8-14~\tev$ LHC can compete with Super-Kamiokande, but a $100~\tev$ hadron collider with reasonable luminosity could overcome the tremendous luminosity advantage of Super-Kamiokande and provide far greater sensitivity.
The scaling of the relevant process rates with luminosity and energy imply
that these results are very robust against even order-of-magnitude changes in the effective luminosity, backgrounds,
or in the number of events needed for a statistically significant measurement.  However, uncertainties in the hadronic
matrix elements can have a large effect on the sensitivity of rare event searches because the event rate depends on
these matrix elements to a high power.

Of these search strategies, constraints from
Super-Kamiokande currently appear to be the most stringent, limiting the mass scale of new physics to respect
$\Lambda_Q > 1.6~\tev$. While there are interesting LHC prospects for $B/L$-processes involving charm and strange quarks, the
LHC sensitivity reach to processes involving first generation quarks ($\Lambda_Q < {\cal O}(2000)~\gev$) signifies that the upgraded LHC may
have enough energy to complement Super-Kamiokande.
But a future ${\cal O}(100~\tev)$ hadron collider could well supersede the Super-Kamiokande
bound and permit production of colored particles which mediate a proton annihilation interaction.  In the case of
either the LHC or a future higher-energy collider, it is
likely that the most successful strategy would be through the direct production of new mediators, but it would also be possible to surpass future Kamioka searches for new physics through indirect processes in which virtual heavy particles mediate rare interactions.

\acknowledgements
We thank Andrew Long, Adam Martin, Hiren Patel, and Surjeet Rajendran for useful discussions.  JB thanks the theory division at
CERN for hospitality while portions of this work were completed.
JGL thanks KITP at UCSB for its hospitality (supported in part by the
National Science Foundation under Grant No.~NSF PHY11-25915) during the writing of this paper.
The work of JK and JGL is supported in part by DOE grant DE-SC0010504.

\appendix

\section{Proton annihilation collider analysis details}
\label{app:collider}

The quark-level $B+L$ violating operator of Eq.~\eqref{eq:Oq1} were implemented in FeynRules 2 \cite{Alloul:2013bka}, and events were
generated in MG5AMC@NLO \cite{Alwall:2014hca}. All partons were required to have a rapidity $|\eta| < 2.5$ and an inter-parton separation $\Delta R < 0.4$. Complexities introduced by an eight-fermion operator with identical final state fermions
and a nontrivial color structure prohibited hadronic showering\footnote{While Pythia 8 \cite{Sjostrand:2007gs} has improved on Pythia 6.4 \cite{Sjostrand:2006za} and can shower events with non-standard color structures, it is so far limited to showering $2 \to 3$ scattering processes.} and detector simulation.  Instead, the parton-level cross-section was calculated, and collider effects were estimated  using hadron and lepton acceptance efficiency curves published for the ``SR04" bin of Ref.~\cite{Chatrchyan:2013fea}.

The presence of eight fermions, some of these identical, presents a computational ambiguity (namely, the point in the calculation when one should apply additional necessary combinatoric factors) that is as yet unresolved in the FeynRules to MG5AMC@NLO simulation chain for processes with more than four identical fermions and fermion-flow violation. Therefore, to study the operators of Eq.~\eqref{eq:Oq1}, another effective operator with the same kinematic structure was employed,
\begin{align}
{\cal O}_{Q1, {\rm standin}} =& {24 \over \Lambda_{Q1}^8} (\bar Q P_L Q)^2 (\bar u_R P_R e_R) (\bar e_R P_R u_R) + {\rm h.c.} \label{eq:spoofop}
\end{align}
which does not violate fermion-flow. The additional factor of twenty-four accounts for the combinatoric difference in contracting the operator of Eq.~\eqref{eq:Oq1} vs. \eqref{eq:spoofop}. Because the $\ell^\pm ~\ell^\pm  + 4 ~{\rm jets}$ search is insensitive to the charge of light quark jets, for this study it is then sufficient to use the operators in Eq.~\eqref{eq:spoofop} to calculate numerical cross-sections of e.g. $uu \to uu d \bar{d} e^+ e^-$. Note that ${\cal O}_{Q1, {\rm standin}}$ producing opposite-sign dileptons is not problematic, it merely requires that our analysis count opposite-sign dilepton events produced by ${\cal O}_{Q1, {\rm standin}}$ as same-sign dilepton events.

After creating parton-level events with MadGraph5 aMC@NLO, we used the fitted efficiency formulae of Ref.~\cite{Chatrchyan:2013fea} to determine collider acceptance of events. For the sum of jet $H_T$ and the jets themselves, the relevant efficiency formula is
\begin{align}
\epsilon_{H_T,j} (p_T) = 0.5 \times \left({\rm Gerf} \left[\frac{p_T - x_{1/2}}{\sigma_p}\right] + 1 \right),
\label{eq:eff1}
\end{align}
where Gerf is the Gaussian error function and $(\epsilon_{\infty},x_{1/2},\sigma_p)$ are fit parameters whose values are given in Table \ref{tab:eff}.
\begin{table}[h!]
\begin{center}
\begin{tabular}{c}
\begin{tabular}{|l||r|r|}
\hline
Eff. variables, $H_T$ and jets &  $x_{1/2}$ (GeV) & $\sigma_p$ (GeV) \\
\hline
\hline
$H_T >$ 400 GeV & 378 & 59.4 \\
\hline
jet w/$p_T > 40$ GeV & 30 &  19 \\
\hline
\end{tabular}
\\
\\
\begin{tabular}{|l||r|r|r|}
\hline
Eff. variables, $\mu$s and $e$s &  $\epsilon_{\infty}$ & $\epsilon_{10}$ & $\sigma_p$ (GeV) \\
\hline
\hline
Electron  & 0.64 & 0.17 & 37\\
\hline
Muon  & 0.67 &  0.33 & 30\\
\hline
\end{tabular}
\end{tabular}
\end{center}
\caption{This table catalogues fitted efficiency parameters used in Eqs.~\eqref{eq:eff1} and \eqref{eq:eff2} to determine acceptance of jets, electrons, muons, and total scalar sum of jet transverse momenta ($H_T$).}
\label{tab:eff}
\end{table}
The efficiency formula for acceptance of muons and electrons is given by
\begin{align}
\epsilon_{e,\mu} (p_T) = \epsilon_{\infty} \times {\rm Gerf} \left[\frac{p_T - 10}{\sigma_p}\right] + \epsilon_{10} \times \left(1- {\rm Gerf} \left[\frac{p_T - 10}{\sigma_p}\right] \right),
\label{eq:eff2}
\end{align}
where again the values for these parameters are displayed in Table \ref{tab:eff}. Finally, the jets and same-sign dilepton signal region we employ in this study requires missing transverse energy $MET > 50~\gev$. Of course, the events we generate have no missing transverse energy at parton level. The efficiency for finding $50~\gev$ of missing energy in this channel for an event with no missing energy, as reported in \cite{Chatrchyan:2013fea}, is $\sim 20 \%$.

\bibliographystyle{JHEP}

\bibliography{gutevap}

\end{document}